\documentclass{aa}  
\usepackage{lscape}
\usepackage{graphicx}
\usepackage{txfonts}
\usepackage[figuresright]{rotating}
\begin{document}
\title{Stars and brown dwarfs in the $\sigma$~Orionis cluster. II}   
\titlerunning{A proper motion study in $\sigma$~Orionis}
\subtitle{A proper motion study}
\author{Jos\'e A. Caballero}
%
%\offprints{Jos\'e Antonio Caballero, investigador Juan de la Cierva at the UCM
%(\email{caballero@astrax.fis.ucm.es}).}  
%
\institute{
Departamento de Astrof\'{\i}sica y Ciencias de la Atm\'osfera, Facultad de
F\'{\i}sica, Universidad Complutense de Madrid, E-28040 Madrid, Spain.
\email{caballero@astrax.fis.ucm.es}}
\date{Received 22 December 2009 / Accepted 20 January 2010}

% \abstract{}{}{}{}{} 
% 5 {} token are mandatory
\abstract
{} 
{I attempt to fully understand the origin of the stellar and substellar
populations in the young  $\sigma$~Orionis open cluster, which is a benchmark
for star-forming studies. 
Because of the very low proper motion of the cluster, late-type dwarfs with
appreciable proper motion in the foreground of $\sigma$~Orionis can be easily
discarded as targets from expensive spectroscopic follow-up studies.}   
{I use the Aladin sky atlas, USNO-B1, public astrometric catalogues, and
photographic plate digitisations to identify stars with proper motions that are
inconsistent with cluster membership in a circular area of radius
30\,arcmin centred on the early-type multiple system $\sigma$~Ori.
Primarily because of the long time baseline of more than half a century, the
errors in the measured proper motions are lower than 2\,mas\,a$^{-1}$.}  
{Of the 42 stars selected for astrometric follow-up, 37 of them are
proper-motion cluster interlopers.
Some USNO-B1 measurements were affected by partially resolved (visual)
multiplicity and target faintness.
Because of their late spectral types and, hence, red colours, 24 contaminants
had been considered at some point as $\sigma$~Orionis members.
I~discuss how contamination may have affected previous work (especially related
to disc frequencies) and the curious presence of lithium absorption in three
M-dwarf proper motion contaminants.
Finally, I~classify the bright star HD~294297 as a late-F field dwarf unrelated
to the cluster based on a new proper motion measurement.} 
{Although proper motions cannot be used to confirm membership in
$\sigma$~Orionis, they can be instead used to discard a number of cluster member
candidates without spectroscopy.}
\maketitle
%
%________________________________________________________________

\section{Introduction}
\label{introduction}

The \object{$\sigma$~Orionis cluster} ($\tau \sim$ 3\,Ma, $d \sim$ 385\,pc) in
the \object{Ori\,OB\,1b} association is a unique site for investigating the
formation and evolution of stars and substellar objects from several tens of
solar masses to a few Jupiter masses (Garrison 1967; Wolk 1996; B\'ejar
et~al. 1999; Caballero 2007a; Walter et~al. 2008 and references therein). 
Although it is not as young and nearby as other star-forming regions such as the
\object{Orion Nebula Cluster}, \object{$\rho$~Ophiuchi}, or
\object{Chamaeleon~I}+II, $\sigma$~Orionis has the great advantage of being
relatively compact ($\rho_{\rm core} \sim$ 20\,arcmin -- Caballero 2008a) and
having a very low visual extinction (0.04\,mag $< E(B-V) <$ 0.09\,mag -- B\'ejar
et~al. 2004; Sherry et~al. 2008).  
The high spatial density and low extinction do not only facilitate the study of
the initial mass function down to well below the deuterium burning limit
(Zapatero Osorio et~al. 2000; B\'ejar et~al. 2001; Caballero et~al. 2007;
Bihain et~al. 2009), but also the identification and characterisation of young
stars with X-ray emission (Sanz-Forcada et~al. 2004; Franciosini et~al. 2006;
Skinner et~al. 2008; Caballero et~al. 2009)
or discs detected on the basis of their infrared excess (Oliveira et~al. 2006;
Hern\'andez et~al. 2007; Zapatero Osorio et~al. 2007; Luhman et~al. 2008),
hydrogen recombination lines in emission (including H$\alpha$ -- Zapatero Osorio
et~al. 2002; Weaver \& Babcock 2004; Kenyon et~al. 2005; Caballero et~al. 2006;
Gatti et~al. 2008; Fedele et~al. 2009), or jets (Reipurth et~al.~1998; Andrews
et~al. 2004).  

Most of the works listed above compute frequencies of X-ray emitters, H$\alpha$
accretors, or disc hosts, some of which are extensively used in the
literature (e.g., Hern\'andez et~al. 2007). 
Accurate frequency determinations require a precise knowledge of the
$\sigma$~Orionis stellar and substellar populations. 
A large fraction of the cluster member candidates, especially in the inner
20\,arcmin, are known to have features of extreme youth (e.g., very early
spectral types, Li~{\sc i} $\lambda$6707.8\,{\AA} in absorption, H$\alpha$ in
strong broad emission, flux excess in the near- and mid-infrared, spectroscopic
signatures of low gravity -- Caballero 2008c).
However, there is evidence that there is significant contamination by field
dwarfs (Caballero et~al. 2008a; Lodieu et~al. 2009), apart from overlapping
{\em young} stellar populations of the Orion Belt (e.g., Jeffries et~al. 2006;
Caballero 2007a; Sacco et~al. 2007; Gonz\'alez-Hern\'andez et~al. 2008) and
galaxies (Caballero et~al. 2007, 2008b; Hern\'andez et~al. 2007; Caballero
2008c).  

The usual procedure for identifying a true $\sigma$~Orionis member is the
spectroscopic follow-up after photometric selection in a colour-magnitude
diagram.
If no spectroscopy is available and the combination of photometric band passes
is not optimal, then a large number of (foreground) field dwarfs may contaminate
a sample.
One way of maximising the telescope time devoted to the spectroscopic follow-up
of reliable cluster member candidates is to discard beforehand those that have
proper motions inconsistent with cluster membership.
In $\sigma$~Orionis, this method of de-contamination has been applied in only a
few occasions (Caballero 2007a, 2008c; Lodieu et~al. 2009)\footnote{Besides,
Zapatero Osorio et~al. (2008) used the small proper motion of \object{S\,Ori~70}
to indicate that this object is farther away than expected if it were a single
field T dwarf lying in the foreground of $\sigma$~Orionis.}. 
Depending on the accuracy of the data used, stars with proper motions larger
than 10 ({\em Hipparcos} and Tycho-2), 20 (USNO-B1), and 30\,mas\,a$^{-1}$
(2MASS/UKIDSS Galactic Cluster Survey) were classified as foreground field
stars. 
These relatively small values are due to the location of $\sigma$~Orionis close
to the antapex, which combined with a distance of almost 400\,pc lead to a
proper motion of the cluster centre of mass of only ($\mu_\alpha \cos{\delta}$,
$\mu_\delta$) = (+2.2 $\pm$ 1.2, --0.5 $\pm$ 1.0)\,mas\,a$^{-1}$ (Caballero
2007a); this proper motion is consistent with the value provided by Kharchenko
et~al.~(2005).  

In this work, I~use Virtual Observatory tools and data archives to search for
and characterise field stars in the direction of $\sigma$~Orionis with 
relatively large proper motions, measured with accuracies superior to
2\,mas\,a$^{-1}$. 
Most of the interloper stars seem to be of late spectral types. 
Because of the resemblance between their magnitudes and colours and those of
young late-type stars in $\sigma$~Orionis, an important fraction of the
proper-motion interlopers had actually been selected as cluster member
candidates in photometric surveys (Scholz \& Eisl\"offel 2004; Sherry et~al.
2004; Kenyon et~al. 2005; Caballero 2006; Hern\'andez et~al. 2007).

\section{Analysis}
\label{analysis}

\subsection{Aladin search}
\label{section.aladinsearch}

I used the stellar proper motions tabulated by the United States Naval
Observatory USNO-B1 catalogue (Monet et~al. 2003), which has a wider coverage of
target magnitudes and is less affected by unresolved binarity than the most
recent Positions and Proper Motions-Extended catalogue (PPMX -- R\"oser
et~al. 2008). 
The Lick Northern Proper Motion 2 catalogue (LNPM2 -- Hanson et~al. 2004), which
also covers the Orion region, is severely affected by
systematics\footnote{For example, all LNPM2 stars in the survey area except two
have very low proper motions. 
The two exceptions, with tabulated LNPM2 proper motions of about
10\,mas\,a$^{-1}$, are an anonymous binary with true $\mu \sim 0$ and G~99--20,
a high proper-motion star with true $\mu \sim$ 290\,mas\,a$^{-1}$ (see below).}. 

\begin{figure}
\centering
\includegraphics[width=0.49\textwidth]{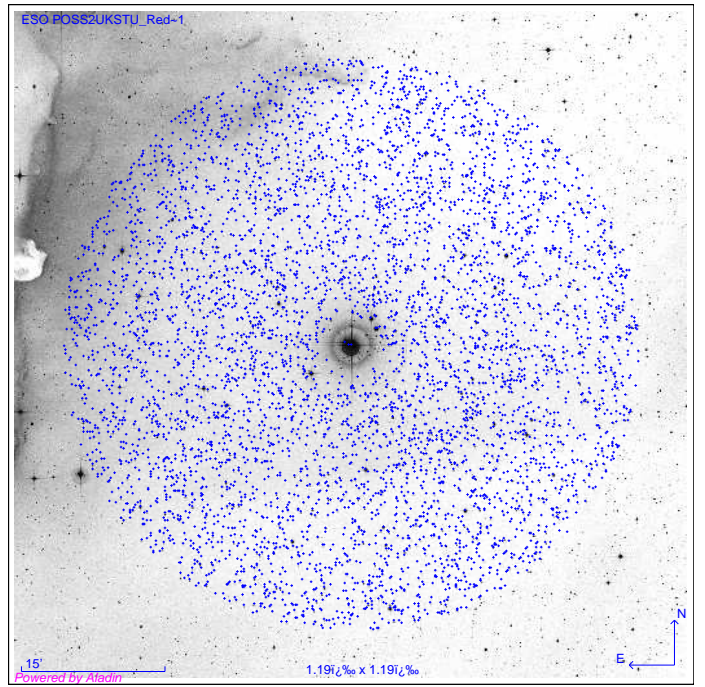}
\caption{Inverse-colour digitised image of the red plate ($R_J$) of the Digital
Sky Survey II centred on the Trapezium-like $\sigma$~Ori system, as shown by
Aladin. 
Cross-matched USNO-B1/2MASS sources are marked with small (blue) crosses.
Size is 70\,$\times$\,70\,arcmin$^2$, north is up, and east is to the left.
Note a part of the Horsehead Nebula (Barnard~33) and the IC~434 reflection
nebula to the east.}   
\label{fu_aladin}
% Aladin
\end{figure}

First, I used the Aladin sky atlas (Bonnarel et~al 2000) to cross-match the
USNO-B1 and Two-Micron All Sky Survey (2MASS -- Skrutskie et~al. 2006)
catalogues in a circle area of radius $\rho$ = 30\,arcmin centred on the
Trapezium-like multiple stellar system $\sigma$~Ori (Fig.~\ref{fu_aladin}),
which is located at the centre of the $\sigma$~Orionis cluster (Caballero
2008b). 
A total of 5421 USNO-B1 sources have a 2MASS near-infrared counterpart within
4\,arcsec (in practice, I~looked for the USNO-B1 counterparts of 2MASS sources).
By cross-matching the two catalogues, it was possible to filter most of the
numerous spurious USNO-B1 sources found in fields with high background and
bright stars, as in this case.
Next, I~used the Aladin tool VOplot to select USNO-B1/2MASS stars with four or
five detections in the USNO-B1 catalogue ($N_{\rm USNO-B1} >$ 3), $J$-band
magnitudes brighter than 15.5\,mag, and proper motions  $\mu_{\rm USNO-B1} >$
40\,mas\,a$^{-1}$. 
Very faint sources or those with detections at only two or three astrometric
epochs (of a maximum of five epochs) have relatively large errors in proper
motion and did not pass the previous filter. 
%$19 >3 && $4 <15.5 && sqrt($17 * $17 + $18 * $18)>40
Of the 5421 USNO-B1/2MASS sources, only 42 (0.8\,\%) objects satisfied my
selection criterion.
They are shown in Table~\ref{table.the42s.mu}.
The USNO-B1 proper motions of the 42 objects vary between 40 and
290\,mas\,a$^{-1}$, with 31 having detections at five epochs. 

%______________________________________________ Figure 

I~chose the 40\,mas\,a$^{-1}$ limit to take into account:
\begin{itemize}
\item the increasing dispersion in intrinsic velocity, which is proportional to
the dispersion in the proper motion, with decreasing masses. 
In a sample of {\em bona~fide} brown dwarfs of the well-known Pleiades cluster,
Bihain et~al. (2006) measured an intrinsic velocity dispersion at least four
times larger than for cluster stars with masses of the order or higher than
1\,$M_\odot$, as expected for a nearly relaxed cluster.
Since limit values of 10\,mas\,a$^{-1}$ had been adopted for the most massive stars
in $\sigma$~Orionis (Caballero 2007a), a limit value four times larger naturally
arises for low-mass cluster stars (although the $\sigma$~Orionis cluster is {\em
not} relaxed -- Caballero 2008b);
\item the increasing number of proper-motion interloper candidates to be
followed-up for decreasing limit values.
While there were 42 objects to be followed-up in the case of the
40\,mas\,a$^{-1}$ limit, there were 77, 161, and 492 such objects for the 30,
20, and 10\,mas\,a$^{-1}$ limits, respectively.
Going below the 40\,mas\,a$^{-1}$ limit would make the target sample
unmanageable; 
\item and the increasing number of USNO-B1 sources with spurious, incorrect,
tabulated values for decreasing proper motions.
\end{itemize}

\subsection{Astrometric follow-up}

I followed the same methodology as Caballero (2009) in measuring high
accuracy proper motions for the 42 objects.
Basically, I~used public data from digitised photographic plates and
astro-photometric catalogues of the last half century.
I~gave special emphasis to the use of data from USNO-A2 (Monet et~al. 1998),
Guide Star Catalogues 2.2 and 2.3 (GSC2.2 and GSC2.3 -- STScI 2001, 2006),
2MASS, Deep Near Infrared Survey of the Southern Sky (DENIS -- Epchtein et~al.
1997), Carlsberg Meridian Catalog 14 (CMC14 -- Evans et~al. 2002; Mui\~nos
2006), UKIRT Infrared Deep Sky Survey (UKIDSS -- Lawrence et~al. 2007), and the
SuperCOSMOS digitisations of plates from the United Kingdom Schmidt Telescope
(UKST) and the first epoch of the Palomar Observatory Sky Survey (POSSI --
Hambly et~al.~2001). 

The relatively poor astrometric precision of the plate digitisations, of
0.4--0.5\,arcsec, was counterbalanced by the long interval between epochs.
In Table~\ref{table.example.epochs}, I~show an example of the nine astrometric
epochs used to compute the proper motion of star No.~24 (G~99--20).
In this case, almost 54 years passed between the first and last epochs.
For the other 41 objects, a typical number of astrometric epochs was also 
eight or nine ($\overline{N} =$ 8.14; with maxima and minima of ten and seven 
epochs, respectively), and they also covered about 54 years. 
Except for the CMC14 epochs, which have an accuracy of one day, all the
tabulated astrometric epochs have a temporal precision better than 0.0001
modified Julian days. 
Depending on when the corresponding field was surveyed, the epochs for each
dataset varied from one target to other, but remained within a narrow interval
(e.g., all 2MASS measurements in the $\sigma$~Orionis area were taken between
J1998 and J2000).
Only the epoch for the the POSSI Red (USNO-A2.0) measurement, J1951.908,
remained identical for all the sources in the area.

%__________________________________________________ Table
   \begin{table}
      \caption[]{Astrometric epochs for star No.~24 (G~99--20).}  
         \label{table.example.epochs}
     $$ 
         \begin{tabular}{ll}
            \hline
            \hline
            \noalign{\smallskip}
Date  		& Image		\\
            \noalign{\smallskip}
            \hline
            \noalign{\smallskip}
1951 Nov 28    	& POSSI Red (USNO-A2.0)	\\
1983 Nov 28    	& Plate 0084 (GSC1.2/ACT)\\
1984 Jan 03    	& UKST Blue (GSC1.2/ACT)\\
1990 Dec 22    	& UKST Red (GSC2.2/2.3)	\\
1996 Jan 27    	& UKST IR		\\
1998 Oct 30    	& 2MASS			\\
1999 Feb 09    	& DENIS			\\
2000 Oct 05    	& CMC14			\\
2005 Oct 06    	& UKIDSS $K$		\\ 
          \noalign{\smallskip}	
            \hline
         \end{tabular}
     $$ 
   \end{table}

When there was no detection by GSC2.2/2.3 and/or USNO-A2.0 (especially because
of target faintness), I~used instead the red optical digitisations of POSSI and
UKST. 
Blue and infrared UKST digitisations were used in all case. 
Star centroid positions were measured with the IRAF environment using standard
tasks. 
Occasionally, there was an additional, intermediate, astrometric epoch at 
$\sim$J1984 from GSC1.2/ACT (Lasker et~al. 1988), or double DENIS detections 
at different epochs between J1995 and J1999.
Apart from this, the faintest objects had in general no CMC14 measurements.
Of the several UKIDSS unmerged astrometric data sets, I~considered those taken
in the $K$ band, which are of higher spatial resolution than to those taken in
the $Y$, $Z$, $J$, and $H$ bands.
The astrometric accuracy for each epoch ranged between 0.06--0.07\,arcsec
for 2MASS (when tabulated, CMC14 accuracy was even higher) to 0.4--0.5 for
USNO-A2.0 and GSC2.2/2.3 or point spread fittings of red targets in the blue
UKST images. 

%______________________________________________ Figure 
\begin{figure}
\centering
\includegraphics[width=0.24\textwidth]{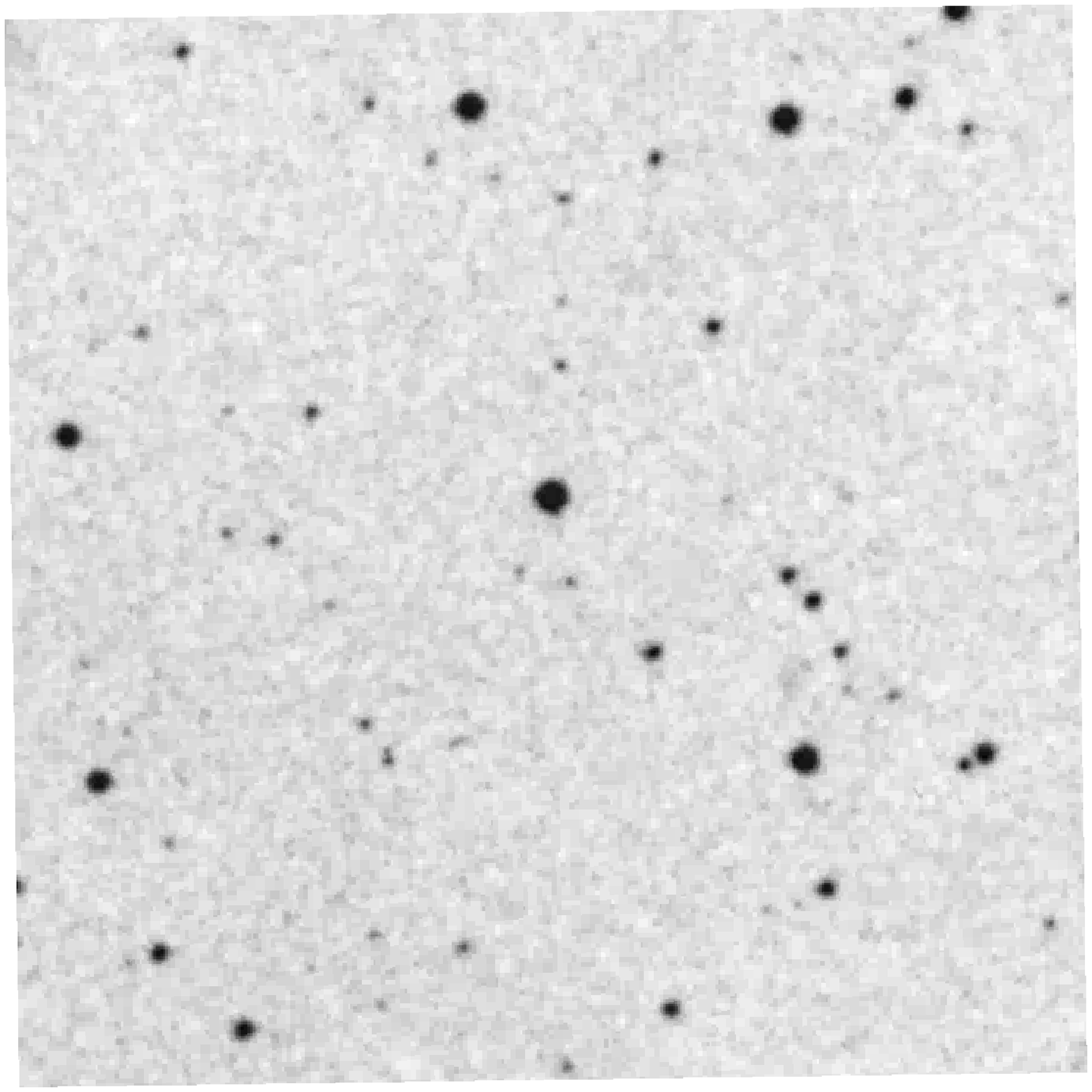}
\includegraphics[width=0.24\textwidth]{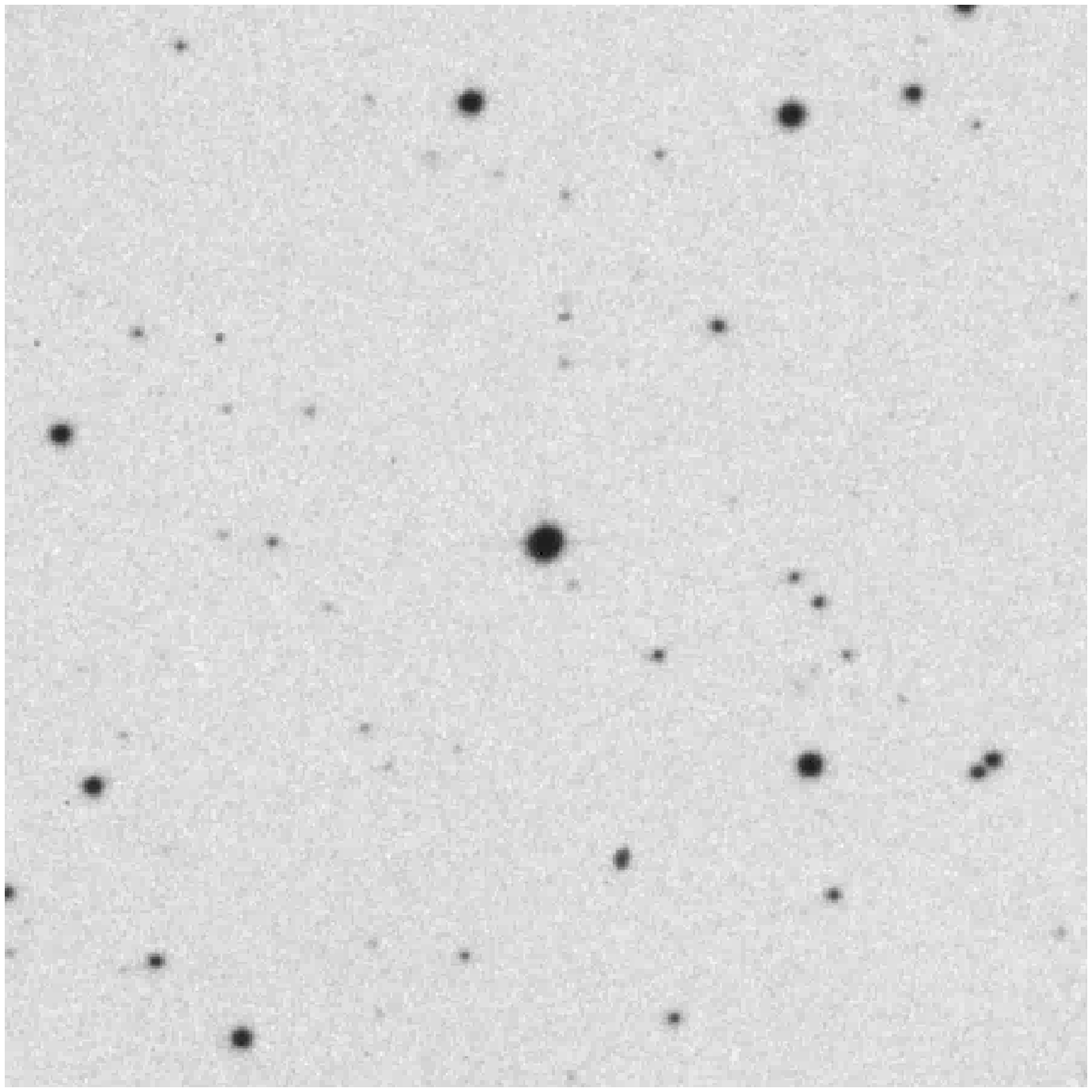}
\caption{Inverse-colour SuperCOSMOS digitised images of the red plates ($R_J$)
of POSSI (J1951.9, {\em left}) and UKST (J1991.0, {\em right}) centred on the
J2000 position of star No.~24 (G~99--20).   
Size is 5\,$\times$\,5\,arcmin$^2$, north is up, and east is to the left.
During the 39 years between the two images, the star moved more than
11\,arcsec.} 
\label{fu_no24ds9}
% DS9 (FollowUp/24)
\end{figure}

\begin{figure}
\centering
\includegraphics[width=0.49\textwidth]{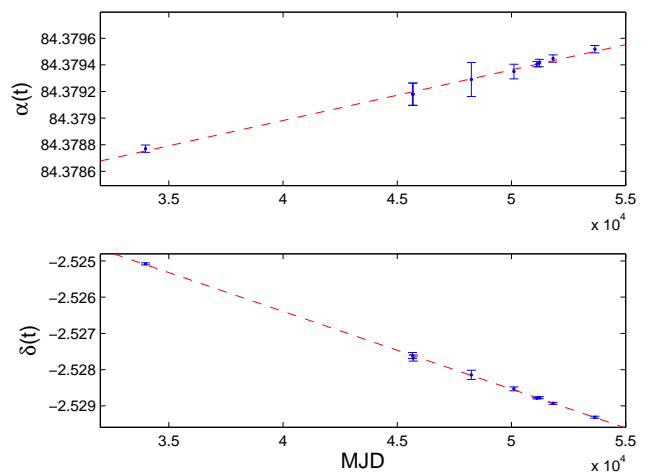}
\caption{Right ascension and declination as a function of modified Julian 
date for the star No.~24 (G~99--20).
Dashed (red) lines indicate the linear fits.
The largest errorbars correspond to the UKST Blue (GSC2.2/2.3) astrometric 
epoch.}   
\label{fu_alphadeltamjd}
% promot (FollowUp/24)
\end{figure}

Proper motions in right ascension $\mu_\alpha \cos{\delta}$ and declination
$\mu_\delta$ were computed independently using simple linear fits with time
(i.e., possible contribution by parallax was not accounted for).
Errors were estimated from the standard deviations of the differences between
the observed values $\alpha(t)$ and $\delta(t)$ and the expected values
$\alpha^*(t)$ and $\delta^*(t)$ from the linear regression.
These errors in proper motion were always smaller than 2\,mas\,a$^{-1}$, with
average proper motion errors of only 0.92 and 0.97\,mas\,a$^{-1}$ in right 
ascension and declination, respectively, which are comparable to the
Tycho-2 catalogue errors (H{\o}g et~al. 2000).
However, the objects studied for this paper are much fainter: only nine of the
42 objects are brighter than $J$ = 12\,mag, while three objects are fainter than
$J$ = 15\,mag.
Because of their red colours, many targets are close to the limiting magnitude
of the blue and red optical photographic plates ($B_J \sim R_F \gtrsim$
20\,mag). 
As examples, Figs.~\ref{fu_no24ds9} and~\ref{fu_alphadeltamjd} show the
temporal variation in the star coordinates and the fit for the high
proper-motion star No.~24 (G~99--20).

The proper motions and number of epochs used in preparing the astrometric
follow-up of the 42 objects are given in the last columns of
Table~\ref{table.the42s.mu}. 

%______________________________________________ Figure 

\section{Results}

\subsection{Objects with incorrect USNO-B1 proper motions}

The correspondence between the USNO-B1 proper motions and those measured in
this work are in general acceptable, with variations smaller than about
10\,mas\,a$^{-1}$, but a number of noticeable exceptions exist.
There are 15 investigated objects whose true proper motions differ from
those inferred by USNO-B1 by 30--120\,mas\,a$^{-1}$.
They were affected by visual multiplicity and faintness in the visible.
The seven cross-matched sources with separations between the USNO-B1 and 2MASS
counterparts of more than 0.9\,arcsec (and up to 3.5\,arcsec) are among those
exceptions (the separations for the other 35 objects are of 0.6\,arcsec or
less).

The remaining 27 stars have reliable USNO-B1 proper motions larger than $\mu >$
40\,mas\,a$^{-1}$.
Of these, six are presented here for the first time (Nos.~01, 03, 19, 25,
28, and~35) and have $J$-band magnitudes and proper motions in the approximate
intervals 40--60\,mas\,a$^{-1}$ and 13--14\,mag, respectively, while the other
21 stars were investigated in previous work with more or less detail.

\subsubsection{Visual multiple systems}
\label{section.binaries}

Ten investigated stars with incorrect USNO-B1 proper motions have visual
companions of roughly the same optical magnitudes at separations $\rho \sim$
4--8\,arcsec, which were incorrectly resolved by at least one USNO-B1
astrometric epoch observation. 
Some properties of the visual companions to the ten stars are shown in
Table~\ref{table.binaries} (stars Nos.~02 and~04 are in fact triple and
quadruple systems, respectively). 
None have proper motions larger than 20\,mas\,a$^{-1}$ or consistent
with membership of a physical system, and only one has ever been considered as a
cluster member candidate (the visual companion to star No.~22; see below).

Three of the ten main targets in visual multiple systems have actual total
proper motions larger than 70\,mas\,a$^{-1}$ (Nos.~02, 12, and~18) and are,
therefore, high proper-motion interloper stars towards the
$\sigma$~Orionis cluster. 
The same is true of another four stars with proper motions $\mu \sim$
20--30\,mas\,a$^{-1}$ (Nos.~09, 20, 21, and~22). 
The remaining three stars (Nos.~04, 38, and~42) have lower proper motions, of
7--12\,mas\,a$^{-1}$, which are consistent with cluster membership.
However, only one of them has ever been considered as a cluster member based on
photometry. 
This star, No.~42 (Mayrit~1610344), was first identified as a (0.39-$M_\odot$,
non-variable) photometric $\sigma$~Orionis member candidate by Scholz \&
Eisl\"offel (2004).  
Afterwards, Sherry et~al. (2004) and Caballero (2008c) agreed on this
classification (although the first authors assigned a membership probability of
only 15\,\%), which should be adhered to until moderate-resolution visible
spectroscopy is obtained. 
The other two low proper-motion stars (Nos.~04 and~38) have
visible/near-infrared (DENIS/2MASS) colours that are inconsistent with cluster
membership for their magnitudes.

Among the seven high proper-motion star contaminants in visual multiple systems,
three targets merit additional descriptions.

\paragraph{No.~18.}
This was first classified as a $\sigma$~Orionis member candidate by Sherry et~al.
(2004), who assigned it an 83\,\% cluster membership probability.
Afterwards, Hern\'andez et~al. (2007) found no infrared flux excess in its
spectral energy distribution, and classified it as a ``class~III'' young object.
However, the star No.~18 has not been identified as a cluster member candidate
in any other photometric survey, including the comprehensive surveys by
Caballero (2008c) and Lodieu et~al. (2009).
Indeed, Sacco et~al. (2008) measured radial velocity and lithium and H$\alpha$
pseudo-equivalent widths that are inconsistent with cluster membership (No.~18
is, in addition, a spectroscopic binary). 
The proper motion measured here, of 78\,mas\,a$^{-1}$, agrees with the Sacco
et~al. (2008) hypothesis that No.~18 is an M2V star in the foreground of
$\sigma$~Orionis. 

\paragraph{No.~21.}
This star has been cited only once, by Hern\'andez et~al. (2007), who classified
it as a ``class~III'' young object. 
However, star No.~21, with a proper motion $\mu \approx$ 31\,mas\,a$^{-1}$ and
red DENIS/2MASS colours, is probably another M dwarf in the foreground of the
cluster. 

\paragraph{No.~22.}
The system is formed by two stars equally separated from an X-ray source detected
with {\em XMM-Newton} by L\'opez-Santiago \& Caballero (2008).
On the one hand, the brightest star in the pair, which has a low proper motion,
has too blue DENIS/2MASS colours for its magnitude, inconsistent with cluster
membership. 
On the other hand, the faintest star (Table~\ref{table.binaries}), which
was also considered to be a cluster member candidate by Sherry et~al. (2004) and
Hern\'andez et~al. (2007), has a proper motion of $\mu \approx$
30\,mas\,a$^{-1}$ and does not belong either to $\sigma$~Orionis. 
X-ray source \object{[LC2008] NX~11} has hardness ratios that are unusual for
young active stars in $\sigma$~Orionis and is located at a relatively large
separation ($\rho \sim$ 4\,arcsec) from the two foreground stars.
The X-ray emission probably originates in an active galaxy that has yet to be
detected by imaging (L\'opez-Santiago \& Caballero 2008).

%__________________________________________________ Table
   \begin{table}
      \caption[]{Companions in visual binary and multiple systems with incorrect USNO-B1 proper motions.}  
         \label{table.binaries}
     $$ 
         \begin{tabular}{ccc ccc}
            \hline
            \hline
            \noalign{\smallskip}
No.  	& $\rho$	& $\theta$	& $\mu_\alpha \cos{\delta}$	& $\mu_\delta$		& $J$ 			\\
  	& [arcsec]	& [deg]		& [mas\,a$^{-1}$]		& [mas\,a$^{-1}$]	& [mag]			\\
            \noalign{\smallskip}
            \hline
            \noalign{\smallskip}
02$^{a}$& 7.2		& 242		&  +3.9$\pm$0.7			&  --2.3$\pm$0.7	& 14.53$\pm$0.03	\\
04$^{b}$& 8.1		& 168		&  +6.1$\pm$0.9			&  --8.1$\pm$1.2	& 16.08$\pm$0.08    	\\
09  	& 5.3		& 182		& --8.2$\pm$1.4			& --13.6$\pm$1.0	& 15.19$\pm$0.05    	\\
12  	& 6.0		& 335		& +10.5$\pm$1.3			&  --7.7$\pm$0.8	& 16.27$\pm$0.08    	\\
18  	& 6.6		& 292		&  +5.4$\pm$1.8 		& --17.2$\pm$2.1	& 14.49$\pm$0.04    	\\
20  	& 4.8		& 234		& --2.0$\pm$1.0 		& --11.5$\pm$1.0	& 15.16$\pm$0.04    	\\
21  	& 7.5		& 152		&  +4.6$\pm$0.7 		& --16.3$\pm$1.1	& 13.07$\pm$0.03    	\\
22  	& 5.6		& 311		&  +1.0$\pm$0.7 		&  --4.5$\pm$0.6	& 14.36$\pm$0.04    	\\
38  	& 6.6		& 324		& --9.6$\pm$0.6			&   +9.7$\pm$0.6	& 14.47$\pm$0.03    	\\
42$^{c}$& 4.5		& 237		&  +1.3$\pm$1.9 		& --10.6$\pm$0.8	& 16.78$\pm$0.03    	\\
          \noalign{\smallskip}	
            \hline
         \end{tabular}
     $$ 
\begin{list}{}{}
\item[$^{a}$] {The visual companion to star No.~02 is appreciably brighter in 
the visible ($\overline{B_J}$ = 18.7\,mag versus $\overline{B_J}$ = 19.8\,mag) 
and fainter in the near infrared ($K_{\rm s}$ = 15.41$\pm$0.08\,mag versus 
$K_{\rm s}$ = 13.59$\pm$0.04\,mag) than the star No.~02.}
\item[$^{b}$] {There is an additional visual companion to star No.~04.
It is a faint, blue source at $\rho \sim$ 4.1\,arcsec, $\theta \sim$ 217\,deg,
only resolved by 2MASS and UKIDSS ($J$ = 16.43$\pm$0.10\,mag), with similarly 
low proper motion.}
\item[$^{c}$] {UKIDSS $J$-band magnitude transformed into the 2MASS one by 
adding the offset $J_{\rm 2MASS} - J_{\rm UKIDSS}$ measured for Mayrit~1610344.
There are two additional visual companions to this star:
USNO-A2.0~0825--01615246, a relatively blue star at 7.8\,arcsec to the southwest
and with a proper motion ($\mu_\alpha \cos{\delta}$,$\mu_\delta$) = 
(+2.9$\pm$1.2,--7.6$\pm$0.7)\,mas\,a$^{-1}$, and a faint red source at about 
3\,arcsec to the southeast only resolved by UKIDSS ($K \approx$ 16.3\,mag).}
\end{list}
   \end{table}

\subsubsection{Single objects}
\label{section.singleobjects}

Discarding the ten previously considered stars in visual multiple systems, there
are another five objects with proper motion differences between USNO-B1 and my
astrometric follow-up of $|\Delta \mu| >$ 12\,mas\,a$^{-1}$.
These variations are ascribed to target faintness\footnote{The large
uncertainties in the proper motions of targets fainter than $J \approx$ 15\,mag
is an {\em a~posteriori} justification of the limit $J <$ 15.5\,mag imposed in
Sect.~\ref{section.aladinsearch}.} (Nos.~08, 23, and~32; the three
of them have $J \gtrsim$ 15\,mag) and high background in the visible (Nos.~29
and~36; both of them have a large separation from the cluster centre and are
embedded in the \object{IC~434} nebula, which is associated with the
\object{Horsehead Nebula} -- Fig.~\ref{fu_aladin}). 
Three of the stars have both colours and magnitudes that are inconsistent with
cluster membership and have never been considered before in photometric surveys.
However, the other two objects, which have variations of $|\Delta \mu| \gg$
12\,mas\,a$^{-1}$, deserve attention.

\paragraph{No.~32 (Mayrit~999306).} 
The faint brown dwarf candidate Mayrit~999306 ([BZR99] S\,Ori~23) was discovered
by B\'ejar et~al. (1999) and has since been identified in several other
independent photometric surveys (B\'ejar et~al. 2001; Gonz\'alez-Garc\'{\i}a
et~al. 2006; Caballero 2008c). 
Its USNO-B1 proper motion, which was based on four astrometric epochs,
was severely affected by its extreme faintness in the blue optical: while
Mayrit~999306 is relatively bright at 2.2\,$\mu$m ($K_{\rm s} \approx$
14.0\,mag), it has photographic magnitudes $B_J \gtrsim R_F \gtrsim$ 20\,mag.
Its true proper motion, of about 14\,mas\,a$^{-1}$ is slightly large, but
still consistent with cluster membership.
Mayrit~999306 still awaits spectroscopic follow-up of moderate spectral
resolution. 

\paragraph{No.~36 (Mayrit~1493050).} 
This young star was discovered by Scholz \& Eisl\"offel (2004).
Afterwards, Kenyon et~al. (2005) found lithium in strong absorbtion and low
gravity features in its spectrum, while Hern\'andez et~al. (2007) claimed that
it harbours an ``evolved disc'' based on IRAC and MIPS {\em Spitzer} photometry.
Maxted et~al. (2008) monitored its radial velocity, which is
consistent with membership and with SB2 binarity. 
Mayrit~1493050 is located close to an overdensity in the IC~434 nebula, which
may lead to an incorrect USNO-B1 measurement, possibly in the $R_F$ plates,
where the nebula is brighter (because of the H$\alpha$ emission).
Its low proper motion, of only 9\,mas\,a$^{-1}$, confirms the membership of
Mayrit~1493050 in $\sigma$~Orionis.

\subsection{Cluster membership and non-membership}

%______________________________________________ Figure 
\begin{figure}
\centering
\includegraphics[width=0.49\textwidth]{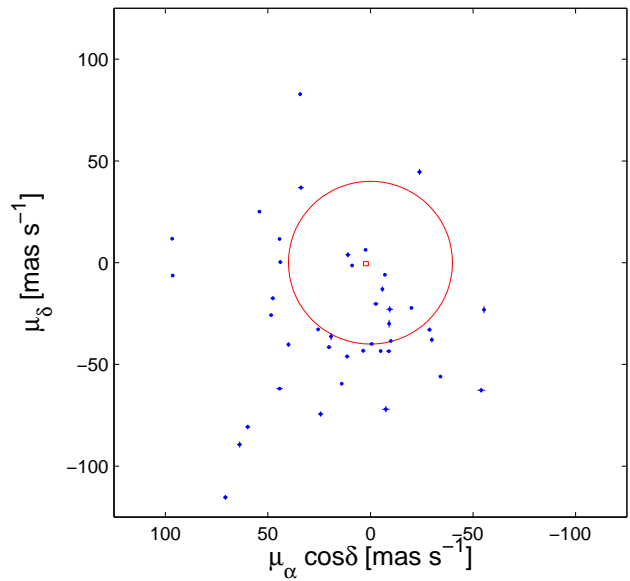}
\caption{Proper motion diagram of the 42 main targets.
Typical errorbars are of the same size as the small (blue) dots.
The (red) circle and small square close to the origin of coordinates
indicate $\mu$ = 40\,mas\,a$^{-1}$ and the proper motion of the $\sigma$~Orionis
cluster centre (Caballero 2007a), respectively.
Star No.~24 (G~99--20), with a proper motion of almost 300\,mas\,a$^{-1}$, lies
outside the limits of the figure.}   
\label{fu_mudemura}
% Software/mm.m
\end{figure}

Table~\ref{table.the42s.simbad} summarises the membership status of the 42
main targets (see also Fig.~\ref{fu_mudemura}). 
Twenty of them were photometric cluster member candidates in the works by
Scholz \& Eisl\"offel (2004; [SE2004]), Sherry et~al. (2004; [SWW2004]), and
Hern\'andez et~al. (2007; [HHM2007]).
However, only three objects (No.~32/Mayrit~999306, No.~36/Mayrit~1493050 --with
known spectroscopic youth features--, and No.~42/Mayrit~1610344) remain as true
cluster member candiates.
The other 17 cluster member candidates do not belong to $\sigma$~Orionis based
on proper motion measurements and, in some cases, high-quality spectroscopic
information from Kenyon et~al. (2005), Caballero (2006), and Sacco et~al.
(2008).
Of the other 22 stars, 15 had not been investigated in the literature and seven
had been discarded as cluster members by Kenyon et~al. (2005) and Caballero
(2006, 2008c) based on spectroscopy and/or proper motion.

To sum up, I~identify for the first time 21 proper-motion interlopers and
contaminants towards the $\sigma$~Orionis cluster, of which eight were
previously considered as cluster member candidates based on photometry
(indicated by ``Yes?'' in the ``Previous member'' column of
Table~\ref{table.the42s.simbad}). 
Another two stars with low actual proper motion, Nos.~04 and~38, have magnitudes
and colours inconsistent with cluster membership (Sect.~\ref{section.binaries}).
In addition, I~confirm the non-cluster membership of 16 foreground stars, of
which ten had never been discussed before in the literature.
The data for the other six stars with proper motion measurements were tabulated
by Giclas et~al. (1961; No.~24), Caballero (2006; Nos.~15 and~30), and Caballero
(2008c; Nos.~05, 14, and~26).

Finally, of the three objects with low proper motions and suitable magnitudes
and colours for $\sigma$~Orionis membership, two (Nos.~32 and~42, see above)
lack spectroscopy and should be followed-up in a future spectroscopic study.
The third star, No.~36, is the only {\em bona~fide} member of cluster.

\section{Discussion}

\subsection{Comparison to other works}

Of the ten bright Tycho-2 (and {\em Hipparcos}) proper-motion contaminants
towards $\sigma$~Orionis identified by Caballero (2007a), only two had $\mu >$
40\,mas\,a$^{-1}$ (\object{HD~294269}, a G0 star with lithium abundance and
radial velocity inconsistent with cluster membership, and
\object{TYC~4770~924~1})\footnote{The same can be applied to the non-Tycho-2
star \object{HD~294276}, which was recovered from the Tycho-1 catalogue.}. 
In spite of the two stars being cross-matched with counterparts in the
USNO-B1/2MASS search (Sect.~\ref{section.aladinsearch}), having similar
Tycho-2 and USNO-B1 proper motions, and being detected in five USNO-B1
astrometric epochs, they did not pass the VOplot filter. 
It is likely that Tycho-2 stars, with mean epochs of observation 2000.0, are
handled by Aladin/VOplot differently from fainter stars with USNO-B1 proper
motions measured from the digitised plates (they have mean epochs of observation
1970--1980). 
In any case, the work of Caballero (2007a) was devoted to the brightest stars of
$\sigma$~Orionis and is complementary to the study presented here.

The three new stars with $\mu >$ 40\,mas\,a$^{-1}$ in the Mayrit catalogue
(Caballero 2008c) also appear here (Nos.~05, 14, and~26).
The other four non-Tycho-2 proper-motion contaminants without spectroscopy and
$\mu \approx$ 20--37\,mas\,a$^{-1}$ in the Mayrit catalogue did not pass the
VOplot filter. 

Lodieu et~al. (2009) tabulated eight stars with proper motions larger than
30\,mas\,a$^{-1}$, of which only one, No.~16 (with spectroscopic information in
Sacco et~al. 2008), is presented here as a proper-motion interloper.
The other seven stars are:
\begin{itemize}
\item \object{S\,Ori~20}, a faint {\em field} M5.5 dwarf with $N_{\rm
USNO-B1}$ = 3 and with previously insufficient membership information (Barrado y
Navascu\'es et~al. 2003; Kenyon et~al. 2005; Caballero et~al. 2007),
\item three stars of proper motion  $\mu <$ 40\,mas\,a$^{-1}$ (of which one was
a photometric cluster member candidate in Sherry et~al. 2004), and
\item the young binary stars \object{Mayrit~92149~AB} (for which Lodieu et~al.
2009 tabulated the two components) and \object{Mayrit~1106058~AB} (the X-ray
star 2E~1486).
The two of them have true null proper motions: spurious values of almost
400\,mas\,a$^{-1}$ resulted from cross-matching the (unresolved) 2MASS and
(resolved) UKIDSS coordinates\footnote{At least the following $\sigma$~Orionis
binary members with $\rho \sim$ 0.5--2.0\,arcsec should also have very large
spurious proper motion values in a 2MASS/UKIDSS cross-match: 
\object{Mayrit~707162~AB}, \object{Mayrit~1564349~AB} (Caballero 2006), 
\object{Mayrit~1245057~AB}, and \object{Mayrit~1411131~AB} (Caballero 2008c).}.
The two stars were first presented as binaries by Caballero (2007b) and
Caballero (2008c), respectively, and exhibit incontrovertible indicators of
youth. 
\end{itemize}

\subsection{Interloper M dwarfs with lithium?}

Following Kenyon et~al. (2005), ``factors of 100 depletion in lithium are
probably needed to reduce'' values of Li~{\sc i} pseudo-equivalent widths, 
pEW(Li~{\sc i})s, to about 300\,m{\AA} or weaker in the spectra of late-type
objects. 
This is because the lithium is burned in the fully convective interiors of
low-mass pre-main sequence stars during the first 20--140\,Ma.
However, the typical pEW(Li~{\sc i}) of T~Tauri stars in $\sigma$~Orionis
ranges between 500 and 600\,m{\AA}, as expected for $\sim$\,3\,Ma-old objects 
(Zapatero Osorio et~al. 2002; Kenyon et~al. 2005; Maxted et~al. 2008; Sacco
et~al. 2008 -- but there might be a spread in lithium abundance).

Three of the proper-motion contaminants in this work have high signal-to-noise
ratio spectroscopy around H$\alpha$ and Li~{\sc i}, taken by Sacco et~al. (2008)
with FLAMES at the VLT.
They are the stars Nos.~11, 16, and~18, which have radial velocities and
pEW(Li~{\sc i})s (and H$\alpha$ emission?) inconsistent with membership in
$\sigma$~Orionis (Table~\ref{table.lithiumdepleted}).   
Their proper motions vary between 60 and 100\,mas\,a$^{-1}$, which are
values much greater than expected for any cluster member.
However, in spite of their late spectral type (M1.5--3.0), the spectra of these
{\em field} dwarfs apparently display lithium in absorption, as measured with
exquisite accuracy by Sacco et~al. (2008).
The pEW(Li~{\sc i})s, of $\sim$\,80--100\,{\AA}, are more than five times
lower than typical $\sigma$~Orionis members of the similar spectral types and
magnitudes, but still appreciable.
If the pEW(Li~{\sc i})s were correct, then the three M dwarfs should be of
the age of the Pleiades or younger ($\tau \lesssim$ 120\,Ma).
I~instead consider the feature at the Li~{\sc i} wavelength that Sacco
et~al. (2008) found was not lithium in absorption, but a collection of lines of
molecular CN ($^{12}$C$^{14}$N -- Grevesse 1968; Mandell et~al. 2004; Ghezzi
et~al. 2008).
The later the spectral type of a star is, the stronger the molecular bands of
lines become.
In M1--3V stars, the CN features may be strong enough to contaminate the lithium
region.
This CN contamination might also affect the spectra of three young
$\sigma$~Orionis member candidates reported by Sacco et~al. (2007) that
exhibited lithium depletion at the level of the three field M dwarfs in
Table~\ref{table.lithiumdepleted}. 

%__________________________________________________ Table
   \begin{table}
      \caption[]{Interloper M dwarfs with supposed lithium in absorption.}
         \label{table.lithiumdepleted}
     $$ 
         \begin{tabular}{cc lcc}
            \hline
            \hline
            \noalign{\smallskip}
No.  	& [SFR2008]	& Sp.	& pEW(Li~{\sc i})	& V$_r$		\\
  	&		& type	& [m\AA]		& [km\,s$^{-1}$]\\
            \noalign{\smallskip}
            \hline
            \noalign{\smallskip}
11  	& S54		& M1.5	& 88$\pm$7		& +41.6$\pm$0.6	\\
16  	& S38		& M3.0	& 78$\pm$5		&  +3.5$\pm$0.5	\\
18  	& S50		& M2.0	& 94$\pm$5		&  +6$\pm$3	\\
          \noalign{\smallskip}	
            \hline
         \end{tabular}
     $$ 
   \end{table}

\subsection{HD~294297}

For completeness, I~also measured the proper motion of the late-F/early-G star
\object{HD~294297}.
Identified as a photometric {\em non-member} in the Ori\,OB\,1b association
subgroup in the early work of Warren \& Hesser (1978), the measurement of a
high lithium abundance in HD~294297 by Cunha et~al. (1995),
$\log{\epsilon({\rm Li})_{\rm NLTE}}$ = 2.56, was in contrast used as an
evidence of its youth. 
Furthermore, the star was one of the only two ``young solar-type members of the
Orion association'' that were observed with STIS onboard the {\em Hubble Space
Telescope} to investigate their boron abundance (Cunha et~al. 2000).
However, Caballero (2007a), who retrieved an abnormally high proper motion from
the Tycho-1 catalogue (but with generous errorbars), again casted doubts on its
membership in Ori\,OB\,1b.
Afterwards, Gonz\'alez~Hern\'andez et~al. (2008) derived a lithium abundance
similar to that measured by Cunha et~al. (1995), $\log{\epsilon({\rm Li})_{\rm
NLTE}}$ = 2.75$\pm$0.18, but using a warmer effective temperature of
6450$\pm$100\,K (instead of 6150\,K). 
In contrast to what is observed in the HD~294297 spectra, stars of this
effective temperature barely show evidence of lithium destruction at the age of
the Pleiades or even older (e.g., Soderblom et~al. 1993). 
Gonz\'alez~Hern\'andez et~al. (2008) found that HD~294297 is
overluminous in the $\mathcal{L}$ versus $J$ diagram with respect to
$\sigma$~Orionis members, where $\mathcal{L}$ is the luminosity per unit of
mass, normalised to this quantity in the Sun.
D'Orazi et~al. (2008) presented a FLAME spectrum of HD~294297, which may
help ascertain the nature of the star, but they did not discuss their results.

In this work, I~was able to use five astrometric epochs from UKIDSS, CMC14,
2MASS, USNO-A2, and the Astrographic Catalogue AC2000.2 (Urban et~al. 1998).
Since the AC2000.2 epoch of observation of HD~294297 was J1895.475, then the 
time interval between the first and last epochs was longer than 110 years. 
The measured proper motion, of ($\mu_\alpha \cos{\delta}$,
$\mu_\delta$) = (+24.9 $\pm$ 1.4, --24.2 $\pm$ 1.1)\,mas\,a$^{-1}$, is similar
to those tabulated in PPMX and UCAC3 (Zacharias et~al. 2009)\footnote{Cunha
et~al. (1995) reported a proper motion ($\mu_\alpha \cos{\delta}$,
$\mu_\delta$) = (+10.5, --1.8)\,mas\,a$^{-1}$.}, which were not available in
2007.
The proper motion is inconsistent with membership in any young association
subgroup in Orion.
HD~294297 is probably a late-F field dwarf at a different heliocentric distance
than Ori\,OB\,1b.

\subsection{Caveats}

%__________________________________________________ Table
   \begin{table*}
      \caption[]{SIMBAD stars with USNO-B1 proper motions between 30 and 40\,mas\,a$^{-1}$.}  
         \label{table.mu30}
     $$ 
         \begin{tabular}{l cc cc c l}
            \hline
            \hline
            \noalign{\smallskip}
Name  						& $\alpha^{J2000}$	& $\theta^{J2000}$	& $\mu_\alpha \cos{\delta}$	& $\mu_\delta$		& $J$ 			& Remarks$^{a}$		\\
  						& 			& 			& [mas\,a$^{-1}$]		& [mas\,a$^{-1}$]	& [mag]			&			\\
            \noalign{\smallskip}
            \hline
            \noalign{\smallskip}
\object{[SWW2004] J053856.630-025702.20}	& 084.735971		& --02.950597		& +28				& --24			& 12.10$\pm$0.03    	& 				\\
\object{[HHM2007] 752}				& 084.706060		& --02.832448		& +16				& --36			& 13.44$\pm$0.03    	& 				\\
\object{[KJN2005] 4.03~29}			& 084.595128		& --02.758452		& --22				& --30			& 14.32$\pm$0.03    	& No V$_r$ in Bu05		\\
\object{[HHM2007] 527}				& 084.616111		& --02.699823		& +30				& +4			& 14.66$\pm$0.04    	& 				\\
\object{2MASS J05392675--0233378}		& 084.861494		& --02.560502 		& +34				& --14			& 11.44$\pm$0.03    	& No $\mu$ in Ca08c		\\
\object{2MASS J05365466--0232069}		& 084.227779 		& --02.535264		& --18				& +26			& 11.50$\pm$0.02    	& No Li~{\sc i} and no $\mu$ in Ca06\\ % SO221028
\object{2MASS J05381906--0232014}		& 084.579439		& --02.533739		& --28				& +16			& 11.99$\pm$0.03    	& No Li~{\sc i} in Ca06		\\ % SO121219
\object{[HHM2007] 120}				& 084.406798		& --02.428913		& --4				& --32			& 14.66$\pm$0.05    	& 				\\
\object{[HHM2007] 1242}				& 084.962323		& --02.397435		& --38				& --10			& 14.46$\pm$0.04    	& 				\\
\object{[HHM2007] 998}				& 084.833748		& --02.365726		& --34				& --8			& 15.08$\pm$0.03    	& 				\\
          \noalign{\smallskip}	
            \hline
         \end{tabular}
     $$ 
\begin{list}{}{}
\item[$^{a}$] {``No V$_r$'', ``no Li~{\sc i}'', and ``no $\mu$'' stand for 
radial velocity, pEW(Li~{\sc i}), and proper motion inconsistent with cluster 
membership, respectively.
The references are Burningham et~al. (2005), Bu05, Caballero (2006), Ca06, and
Caballero (2008c), Ca08c.}
\end{list}
   \end{table*}
%
% It comes from \subsection{Going further in proper motion}   
%

In Sect.~\ref{introduction}, I~noted the importance of a correct
de-contamination before deriving any typical parameter in studies of
star-forming regions.
For example, special care has been applied when measuring the shape of the
initial mass function in $\sigma$~Orionis (see extensive de-contamination
discussion in, e.g., Caballero et~al. 2007, Lodieu et~al. 2009, and Bihain et~al.
2009). 
However, I~warn against the incorrect use of certain parameters that were
derived without a correct de-contamination:

\paragraph{Total number of stars.}
Some authors, such as Sherry et~al. (2004) and Walter et~al. (2008), 
estimated that the total numbers of stars and brown dwarfs in $\sigma$~Orionis
is of about 600--700 based on visible surveys without near-infrared follow-up. 
However, the latest comprehensive analyses (e.g., Caballero 2008c; Lodieu et~al.
2009) indicate that there are no more than 400 cluster objects above the
deuterium burning mass limit. 
The reader must not misinterpret this information and automatically apply a
one-half correction factor to the works of the New York group.
Part of the problem arose from incorrect assumptions about the slopes of the
mass function in different mass regimes and the size of the $\sigma$~Orionis cluster:
the actual contamination degree can be lower than one third. 
Only ten proper-motion contaminants (about 25\,\% of this full
sample) were classified as cluster member candidates in the $(B)VRI$ survey of
Sherry et~al. (2004).

\paragraph{Frequency of discs.}
According to Scholz \& Eisl\"offel (2004), 5--7\,\% of $\sigma$~Orionis stars
and brown dwarfs in the mass range 0.03 to 0.7\,$M_\odot$ possess a
circumstellar disc.
This percentage interval contrasts with the values compiled by Caballero (2007a)
and the widely used value of $\sim$35\,\% of Hern\'andez et~al. (2007) for
$\sigma$~Orionis stars in the mass range 0.1 to 1.0\,$M_\odot$.
However, the contamination in the works by Scholz \& Eisl\"offel (2004) and
Hern\'andez et~al. (2007) is significant.
On the one hand, Hern\'andez et~al. (2007) considered ten proper-motion
contaminants in this study to be cluster members, including the Gliese et~al.
(1961) high proper-motion star G~99--20; proper-motion contaminants may be
a small fraction of all the contaminants in a photometric survey. 
On the other hand, Scholz \& Eisl\"offel (2004) considered only five
proper-motion contaminants, but covered only about one third of the area of
the $\sigma$~Orionis cluster.
Most of their member candidates lacked spectroscopic information.
More precisely determined frequencies of discs for different mass intervals, at
about 40--60\,\%, can be found in Luhman et~al. (2008; who used the Mayrit
catalogue), Caballero et~al. (2007), and Zapatero Osorio et~al. (2007) for
stars, brown dwarfs, and planetary-mass objects, respectively.

\subsection{Going further in terms of proper motion}

In Sect.~\ref{section.aladinsearch}, I~noted that there were 33 USNO-B1 sources
within the survey area that have proper motions between 30 and
40\,mas\,a$^{-1}$ (and 119 between 20 and 40\,mas\,a$^{-1}$, and 450 between 10
and 40\,mas\,a$^{-1}$). 
With Aladin, I~searched for the SIMBAD counterparts of the 33 candidate
interlopers.
Ten of them had been discussed in the literature (Table~\ref{table.mu30}), 
four having been tabulated already as interlopers in the Mayrit catalogue (three of
the four have spectroscopy).
The other six interloper candidates were likely contaminants in the works of
Sherry et~al. (2004 -- one interloper) and Hern\'andez et~al. (2007 -- five
interlopers), which casts additional doubt on, e.g., the frequencies of discs
derived in those studies. 
We clearly require a massive astrometric follow-up of as many as possible
$\sigma$~Orionis member candidates that have no known features of youth.

\section{Conclusions}

I have used the Aladin Virtual Observatory tool to search for proper-motion
contaminants towards the $\sigma$~Orionis cluster.
Of the 5421 USNO-B1/2MASS cross-matched sources in an area of radius
30\,arcmin centred on the Trapezium-like star of eponymous name, only 42 of them
had USNO-B1 proper motions larger than 40\,mas\,a$^{-1}$, 2MASS $J$-band
magnitudes brighter than 15.5\,mag, and more than three USNO-B1 detections.
I~have astrometrically followed them up using a number of catalogues (e.g.,
USNO-A2, GSC2.2/2.3, 2MASS, DENIS, CMC14, UKIDSS) and photographic plate
digitisations from SuperCOSMOS, and measured accurate proper motions with
errorbars of less than 2\,mas\,a$^{-1}$.
The USNO-B1 proper motions of several targets were affected by partially
resolved (visual) multiplicity or faintness in the visible.

Of the 42 investigated objects, 27 had previously been considered to be
$\sigma$~Orionis member candidates.  
The other 15 objects are reported here for the first time.
I~divide the 42 objects into:
\begin{itemize}
\item one {\em bona~fide} cluster member of low actual proper motion ($\mu <$
10\,mas\,a$^{-1}$) with known spectroscopic features of youth
(No.~36/Mayrit~1493050); 
\item two cluster member candidates of low proper motion, including a brown
dwarf (No.~32/Mayrit~999306 and No.~42/Mayrit~1610344);
\item two previously unknown stars of low proper motion with magnitudes and
colours inconsistent with cluster membership (Nos.~04 and~38);
\item 37 proper-motion interlopers, of which 21 are firstly discarded for
cluster membership here.
In a few cases, the proper-motion measurement is supported by spectroscopic
information (e.g., lack of lithium in absorption) obtained in other works.
\end{itemize}

Interestingly, 17 of the 37 proper-motion interlopers (46\,\%) were
tabulated by Scholz \& Eisl\"offel (2004), Sherry et~al. (2004), and Hern\'andez
et~al. (2007) as $\sigma$~Orionis member candidates, although the membership of
a few of them were subsequently discarded by Kenyon et~al. (2005),
Caballero (2006, 2008c), or Sacco et~al. (2008).
The proper-motion interlopers in this work outnumber those in other
comparable studies, such as Caballero (2008c) or Lodieu et~al. (2009). 

I~discussed the curious resemblance between the pseudo-equivalent widths of
the Li~{\sc i} $\lambda$6707.8\,{\AA} line in absoprtion of three early-M
proper-motion contaminants and three ``lithium-depleted young star'' candidates.
I~also measured the proper motion of the star HD~294297 with a time baseline of
more than 110 years and concluded that it is not a young F/G star, but a
late-F field dwarf unrelated to Ori\,OB\,1b.
Finally, I~showed some examples of studies in which de-contamination (by
proper-motion interlopers) was not correctly accomplished, and the
consequent implications of their results.
Preliminary analyses of stars with USNO-B1 proper motions between 30 and
40\,mas\,a$^{-1}$ indicate that contamination by foreground interlopers could
severely affect the results of some widely used works.

This analysis is a new, but necessary, step to characterise in an optimal
way the stellar and substellar populations of the $\sigma$~Orionis cluster,
which is one of the most accesible laboratories of star formation.

\begin{acknowledgements}

I am grateful to the anonymous referee for his/her fast and considerate
report. 
I thank J.~I. Gonz\'alez Hern\'andez for helpful discussion on lithium
abundance. 
Formerly, I~was an investigador Juan de la Cierva at the Universidad
Complutense de Madrid; 
currently, I~am an investigador Ram\'on y Cajal at the Centro de
Astrobiolog\'{\i}a (CSIC-INTA).
This research has made use of the SIMBAD, operated at Centre de Donn\'ees
astronomiques de Strasbourg, France, and the NASA's Astrophysics Data System.
Financial support was provided by the Universidad Complutense de Madrid,
the Comunidad Aut\'onoma de Madrid, the Spanish Ministerio Educaci\'on y
Ciencia, and the European Social Fund under grants:
AyA2008-06423-C03-03, 			% (WSO) 
AyA2008-00695,				% (estrellax II)
PRICIT S-0505/ESP-0237,			% (AstroCAM) 
and CSD2006-0070. 			% (Consolider-GTC). 

\end{acknowledgements}

\appendix

\section{Long tables}

{\bf NOTE to the editors: the two Tables~\ref{table.the42s.mu}
and~\ref{table.the42s.simbad} should appear throughout the text, instead of in
the Appendix.}

%__________________________________________________ Table
   \begin{table*}
      \caption[]{USNO-B1 stars towards $\sigma$~Orionis with  
      $\mu_{\rm USNO-B1} >$ 40\,mas\,a$^{-1}$, $J_{\rm 2MASS} <$ 15.5\,mag,
      and $N_{\rm USNO-B1} >$ 3: coordinates, magnitudes, and proper motions.}  
         \label{table.the42s.mu}
     $$ 
         \begin{tabular}{c ccc c ccc c ccc}
            \hline
            \hline
            \noalign{\smallskip}
  					& 
\multicolumn{3}{c}{2MASS}		&
  					& 
\multicolumn{3}{c}{USNO-B1}		&
  					& 
\multicolumn{3}{c}{This work}		\\
        \noalign{\smallskip}
 	\cline{2-4}
 	\cline{6-8}
 	\cline{10-12}
        \noalign{\smallskip}
No.					&
$\alpha^{J2000}$			& 
$\delta^{J2000}$			& 
$J$					& 
					&
$\mu_\alpha \cos{\delta}$		& 
$\mu_\delta$				& 
$N$					&
					&
$\mu_\alpha \cos{\delta}$		& 
$\mu_\delta$				&				
$N$					\\ 
  					& 
					& 
					& 
[mag]					& 
					&
[mas\,a$^{-1}$]				& 
[mas\,a$^{-1}$]				& 
					& 
					&
[mas\,a$^{-1}$]				& 
[mas\,a$^{-1}$]				&
					\\
            \noalign{\smallskip}
            \hline
            \noalign{\smallskip}
01   	& 084.595001	& --03.064613	& 14.59$\pm$0.03	&	& +18   &--40   & 5     &	& +19.2$\pm$1.0     &--36.3$\pm$1.7	& 9	\\ %	   
02	& 084.425714	& --03.025405   & 14.53$\pm$0.03	&	&--82	&--62	& 5     &	&--54.0$\pm$1.9     &--62.7$\pm$1.2	& 8	\\ %	 (***02)	
03   	& 084.826136	& --03.013852   & 13.07$\pm$0.02	&	& +10	&--50	& 5     &	& +11.4$\pm$1.1     &--46.1$\pm$0.9	& 8	\\ %	   
04   	& 084.539224	& --03.003337   & 14.26$\pm$0.03	&	& +30	& +28	& 5     &	& +11.0$\pm$1.2     &  +3.9$\pm$1.3	& 8	\\ %	 (***04)	  
05   	& 084.948355	& --02.983131   & 10.36$\pm$0.03	&	& +42	&--14	& 5     &	& +47.6$\pm$1.1     &--17.5$\pm$0.9	& 9	\\ % J05394761-0258593   
06   	& 084.850070	& --02.963510   & 11.26$\pm$0.03    	&	& +32   & +80   & 5     &	& +34.3$\pm$0.5     & +82.8$\pm$1.0	& 9	\\ % SO411068
07   	& 084.850866	& --02.924599   & 11.79$\pm$0.03    	&	& +14   &--54   & 5     &	& +14.0$\pm$0.7     &--59.5$\pm$0.5	& 9	\\ % SO441143	 
08   	& 084.642299	& --02.859457   & 15.18$\pm$0.04    	&	& +54   &--52   & 4     &	& +44.3$\pm$1.5     &--61.9$\pm$0.7	& 7	\\ %	 (J=15.181mag)   
09	& 084.244956	& --02.819184   & 13.62$\pm$0.04    	&	&  +6   & +70   & 4     &	& --9.4$\pm$1.6     &--22.9$\pm$1.6	& 9	\\ %	 (***09)   
10   	& 084.369333	& --02.779688   & 14.81$\pm$0.03    	&	&--28   &--36   & 5     &	&--29.9$\pm$0.7     &--37.9$\pm$1.2	& 7	\\ % [KJN2005] 38  
11   	& 084.706846	& --02.757468   & 13.15$\pm$0.03    	&	&--54   &--24   & 5     &	&--55.3$\pm$0.9     &--23.1$\pm$1.8	& 8	\\ % [SWW2004] J053849.560-024526.94	
12	& 084.533024	& --02.752464   & 15.33$\pm$0.05    	&	& +20   &--36   & 4     &	& --7.5$\pm$1.8     &--72.0$\pm$1.6	& 7	\\ %	 (***12)   
13   	& 084.485297	& --02.732624   & 13.04$\pm$0.03    	&	& +20   &--40   & 5     &	& +20.2$\pm$1.1     &--41.5$\pm$0.9	& 8	\\ % [SWW2004] J053756.467-024357.36	
14   	& 085.152776	& --02.729698   & 11.54$\pm$0.03    	&	& +50   & +10   & 4     &	& +43.9$\pm$0.9     &  +0.3$\pm$1.0	& 9	\\ % J05403667-0243469  
15   	& 085.098384	& --02.718792   & 11.37$\pm$0.03    	&	& +62   &--96   & 5     &	& +63.8$\pm$1.2     &--89.4$\pm$1.5	& 10	\\ % SO430116	 
16   	& 084.730988	& --02.691593   & 12.17$\pm$0.03    	&	& +96   &  +8   & 5     &	& +96.6$\pm$0.9     & +11.8$\pm$0.7	& 9	\\ % [SWW2004] J053855.425-024129.68, [SFR2008] S38 (***16)	  
17   	& 084.270844	& --02.684343   & 14.37$\pm$0.03    	&	&  +6   &--42   & 5     &	&  +3.5$\pm$1.0     &--43.3$\pm$1.0	& 7	\\ % [SWW2004] J053705.004-024103.43	
18	& 084.523630	& --02.672057   & 12.77$\pm$0.03    	&	&--20   &--40   & 4     &	& +24.3$\pm$1.2     &--74.4$\pm$1.3	& 8	\\ % [SWW2004] J053805.676-024019.36, [SFR2008] S50 (***18)	  
19   	& 085.122689	& --02.651908   & 13.98$\pm$0.03    	&	&--12   &--46   & 5     &	& --8.9$\pm$1.2     &--43.5$\pm$0.5	& 8	\\ %	   
20   	& 084.472224	& --02.602508   & 14.94$\pm$0.05    	&	& +38   & +30   & 5     &	& --2.6$\pm$1.3     &--20.2$\pm$0.9	& 8	\\ %	 (***20)	      
21   	& 085.032746	& --02.598684   & 14.65$\pm$0.04    	&	&  +6   &--40   & 5     &	& --9.1$\pm$0.7     &--30.0$\pm$1.9	& 8	\\ %	 (***21)   
22   	& 084.346086	& --02.546276   & 14.18$\pm$0.04    	&	&--42   &  +0   & 5     &	&--20.0$\pm$0.7     &--22.2$\pm$0.6	& 8	\\ % [SWW2004] J053723.067-023246.38 (***22)	
23   	& 084.257197	& --02.544301   & 15.29$\pm$0.07    	&	& +40   &--18   & 4     &	& +40.0$\pm$0.7     &--40.2$\pm$1.1	& 8	\\ %	 (J=15.293mag)  
24   	& 084.379406	& --02.528789   & 10.46$\pm$0.03    	&	& +50   &--282  & 5     &	& +49.9$\pm$0.4    &--282.0$\pm$0.8	& 9	\\ % G 99-20	
25   	& 085.175498	& --02.516978   & 14.17$\pm$0.03    	&	&--30   &--34   & 5     &	&--28.8$\pm$1.0     &--32.9$\pm$1.0	& 8	\\ %	   
26   	& 084.815180	& --02.499121   &  9.81$\pm$0.03    	&	&  +0   &--42   & 5     &	& --0.6$\pm$1.1     &--39.9$\pm$0.6	& 9	\\ % J05391564-0229568   
27   	& 084.622984	& --02.479908   & 14.87$\pm$0.03    	&	& +32   & +46   & 5     &	& +33.9$\pm$1.4     & +36.9$\pm$1.0	& 8	\\ %	   
28   	& 084.353969	& --02.476976   & 12.75$\pm$0.02    	&	&--10   &--40   & 5     &	& --9.9$\pm$0.4     &--38.5$\pm$0.8	& 8	\\ %	   
29   	& 085.159839	& --02.451720   & 14.02$\pm$0.03    	&	& +78   &--104  & 4     &	& +70.7$\pm$1.1    &--115.3$\pm$1.2	& 7	\\ %	   
30   	& 084.665077	& --02.447131   & 11.22$\pm$0.03    	&	& +46   &--18   & 5     &	& +48.4$\pm$0.9     &--25.8$\pm$0.6	& 10	\\ % SO120908	 
31   	& 084.774245	& --02.437633   & 14.34$\pm$0.03    	&	& --6   &--40   & 5     &	& --5.0$\pm$0.3     &--43.4$\pm$0.6	& 7	\\ % [KJN2005] 30 / [SE2004] 42 
32   	& 084.462951	& --02.435409   & 14.92$\pm$0.04    	&	&+240   &--210  & 4     &	& --5.8$\pm$0.5     &--13.0$\pm$1.3	& 7	\\ % Mayrit 999306 / [BZR99] S Ori 23 (faint; i=17.135mag,J=14.921mag)     
33	& 084.806169	& --02.397745	& 13.95$\pm$0.03    	&	& +50   & +22   & 5     &	& +54.1$\pm$0.5     & +25.1$\pm$0.4	& 8	\\ % [KJN2005] 12
34$^{a}$& 084.362205	& --02.365031	& 11.41$\pm$0.02    	&	&--18   & +44   & 5     &	&--23.9$\pm$1.1     & +44.6$\pm$1.5	& 10	\\ % SO241003 (rho=2.8arcsec,theta=128deg,DZ=2.9mag,DH=3.6mag,no-K)
35	& 084.999408	& --02.339369	& 14.59$\pm$0.03    	&	& +44   & +12   & 5     &	& +44.3$\pm$0.5     & +11.6$\pm$0.6	& 8	\\ %	      
36	& 085.004221	& --02.333267	& 13.10$\pm$0.03    	&	&--12   & +48   & 4     &	& --7.0$\pm$0.6     & --5.9$\pm$0.3	& 7	\\ % Mayrit 1493050 (faint; i=15.307mag,J=13.096mag)	 
37	& 084.923889	& --02.329536	& 14.22$\pm$0.03    	&	& +20   &--36   & 5     &	& +25.5$\pm$0.8     &--32.8$\pm$0.8	& 8	\\ % [SWW2004] J053941.738-021946.22 
38	& 084.720246	& --02.324637	& 14.23$\pm$0.03    	&	& +70   &--108  & 5     &	&  +8.9$\pm$0.7     & --1.4$\pm$0.8	& 8	\\ %	 (***37)  
39	& 084.355084	& --02.267993	& 13.80$\pm$0.04    	&	&--34   &--54   & 5     &	&--34.1$\pm$0.2     &--56.0$\pm$0.3	& 7	\\ % [SWW2004] J053725.220-021604.56
40	& 084.543043	& --02.233226	& 13.47$\pm$0.03    	&	&+104   & --6   & 5     &	& +96.4$\pm$0.4     & --6.3$\pm$0.6	& 8	\\ % [SE2004] 57, [KJN2005] 7	
41$^{b}$& 084.969042	& --02.192280	& 14.85$\pm$0.04    	&	& +54   &--82   & 4     &	& +59.8$\pm$1.2     &--80.7$\pm$1.1	& 8	\\ % [SE2004] 112 (rho=2.4arcsec,theta~177deg,DZ=3.0mag,DKs=2.9mag)	
42$^{c}$& 084.560575	& --02.170903   & 12.48$\pm$0.03    	&	&--52   &--32   & 4     &	&  +2.4$\pm$0.5     &  +6.3$\pm$0.5	& 8	\\ % Mayrit 1610344 / [SE2004] 6 (***41)
          \noalign{\smallskip}	
            \hline
         \end{tabular}
     $$ 
\begin{list}{}{}
\item[$^{a}$] {Star No.~34 is a close binary, only resolved by UKIDSS ($\rho
\sim$ 2.8\,arcsec, $\theta \sim$ 128\,deg, $\Delta Z \sim \Delta K \sim$ 3\,mag).}
\item[$^{b}$] {Star No.~41 is a close binary, only resolved by UKIDSS ($\rho
\sim$ 2.4\,arcsec, $\theta \sim$ 177\,deg, $\Delta Z \sim \Delta K \sim$ 3\,mag).}
\item[$^{c}$] {The USNO-B1 proper motion values of star No.~42 actually
correspond to a faint source at 4.5\,arcsec to the southwest of the tabulated
coordinates (see Sect.~\ref{section.binaries}).}
\end{list}
   \end{table*}
%

%__________________________________________________ Table
   \begin{table*}
      \caption[]{USNO-B1 stars towards $\sigma$~Orionis with  
      $\mu_{\rm USNO-B1} >$ 40\,mas\,a$^{-1}$, $J_{\rm 2MASS} <$ 15.5\,mag,
      and $N_{\rm USNO-B1} >$ 3: alternative names and cluster membership
      status.}   
         \label{table.the42s.simbad}
     $$ 
         \begin{tabular}{c l cccl l l}
            \hline
            \hline
            \noalign{\smallskip}
No.					&
Alternative				&
[SE2004]				&
[SWW2004]				&
[HHM2007]				&
Discarded$^{a}$				&
Previous				&
Current					\\ 
  					& 
name  					& 
  					& 
  					& 
  					& 
  					& 
member$^{b}$  				& 
member$^{b}$				\\
            \noalign{\smallskip}
            \hline
            \noalign{\smallskip}
01      & ...						& ...	& ...	& ...	& {\em This work}	& Unknown	& No	\\ % 
02      & ...						& ...	& ...	& ...	& {\em This work}	& Unknown	& No	\\ % 
03      & ...						& ...	& ...	& ...	& {\em This work}	& Unknown	& No	\\ % 
04      & ...						& ...	& ...	& ...	& {\em This work}	& Unknown	& No	\\ % 
05      & \object{2MASS J05394761--0258593}		& ...	& ...	& ...	& Ca08			& No		& No	\\ % 
06      & \object{SO411068}				& ...	& 202	& ...	& Ca06			& No		& No	\\ % 
07      & \object{SO441143}				& ...	& ...	& ...	& Ca06			& No		& No	\\ % 
08      & ...						& ...	& ...	& ...	& {\em This work}	& Unknown	& No	\\ % 
09      & ...						& ...	& ...	& ...	& {\em This work}	& Unknown	& No	\\ % 
10      & \object{[KJN2005] 38} 			& ...	& ...	& ...	& Ke05			& No		& No	\\ % 
11      & \object{[SWW2004] J053849.560--024526.94}	& ...	& 101	& 753	& Sa08			& No		& No	\\ % 
12      &  ...  					& ...	& ...	& ...	& {\em This work}	& Unknown	& No	\\ % 
13      & \object{[SWW2004] J053756.467-024357.36}	& ...	& 116	& 260	& {\em This work}	& Yes?		& No	\\ % 
14      & \object{2MASS J05403667--0243469}		& ...	& ...	& ...	& Ca08			& No		& No	\\ % 
15      & \object{SO430116}				& ...	& ...	& ...	& Ca06			& No		& No	\\ % 
16      & \object{[SWW2004] J053855.425--024129.68}	& ...	& 71	& 804	& Sa08			& No		& No	\\ % 
17      & \object{[SWW2004] J053705.004--024103.43}	& ...	& 218	& ...	& {\em This work}	& Yes?		& No	\\ % 
18      & \object{[SWW2004] J053805.676--024019.36}	& ...	& 88	& 330	& Sa08			& No		& No	\\ % 
19      &  ...  					& ...	& ...	& ...	& {\em This work}	& Unknown	& No	\\ % 
20      & ...						& ...	& ...	& ...	& {\em This work}	& Unknown	& No	\\ % 
21      & \object{[HHM2007] 1354}			& ...	& ...	& 1354	& {\em This work}	& Yes?		& No	\\ % 
22      & \object{[SWW2004] J053723.067--023246.38}	& ...	& 215	& 27	& {\em This work}	& No?		& No	\\ % 
23      & ...						& ...	& ...	& ...	& {\em This work}	& Unknown	& No	\\ % 
24      & \object{G 99--20}				& ...	& ...	& 74	& Gi61$^{c}$		& No		& No	\\ % 
25      & ...						& ...	& ...	& ...	& {\em This work}	& Unknown	& No	\\ % 
26      & \object{2MASS J05391564--0229568}		& ...	& ...	& 961	& Ca08			& No		& No	\\ % 
27      & \object{[HHM2007] 544}			& ...	& ...	& 544	& {\em This work}	& Yes?		& No	\\ % 
28      & ...						& ...	& ...	& ...	& {\em This work}	& Unknown	& No	\\ % 
29      & ...						& ...	& ...	& ...	& {\em This work}	& Unknown	& No	\\ % 
30      & \object{SO120908}				& ...	& ...	& ...	& Ca06			& No		& No	\\ % 
31      & \object{[SE2004] 42}  			& 42	& 204	& ...	& Ke05			& No		& No	\\ % 
32      & \object{Mayrit 999306}			& ...	& ...	& 209	& ...			& Yes?  	& Yes?  \\ % 
33      & \object{[KJN2005] 12} 			& ...	& ...	& ...	& Ke05$^{d}$		& No		& No	\\ % 
34      & \object{SO241003}				& ...	& ...	& 52	& Ca06			& No		& No	\\ % 
35      & ...						& ...	& ...	& ...	& {\em This work}	& Unknown	& No	\\ %
36      & \object{Mayrit 1493050}			& 122	& ...	& 1323	& ...			& Yes  		& Yes  	\\ % 
37      & \object{[SWW2004] J053941.738--021946.22}	& ...	& 206	& ...	& {\em This work}	& Yes?		& No	\\ % 
38      & ...						& ...	& ...	& ...	& {\em This work}	& Unknown	& No	\\ %
39      & \object{[SWW2004] J053725.220--021604.56}	& ...	& 201	& ...	& {\em This work}	& Yes?		& No	\\ % 
40      & \object{[SE2004] 57}  			& 57	& ...	& ...	& Ke05			& No		& No	\\ % 
41      & \object{[SE2004] 112} 			& 112	& ...	& ...	& {\em This work}	& Yes?		& No	\\ % 
42      & \object{Mayrit 1610344}			& 6	& 228	& ...	& ...			& Yes?  	& Yes?  \\ % 
          \noalign{\smallskip}	
            \hline
         \end{tabular}
     $$ 
\begin{list}{}{}
\item[$^{a}$] {Work in which the cluster membership of the object was firstly
discarded:  
Gi61, Giclas et~al. 1961; Ke05, Kenyon et~al. 2005; Ca06, Caballero 2006; Ca08,
Caballero 2008c; Sa08, Sacco et~al. 2008.}
\item[$^{b}$] {``Yes'': confirmed cluster member with signatures of youth -- 
``Yes?'': cluster member candidate based on photometry -- 
``No?'': probable non-cluster member based on photometry --
``No'': non-cluster member based on spectroscopy, photometry or proper motion --
``Unknown'': firtsly presented in this paper.}
\item[$^{c}$] {Actually, in 1961, the existence of the $\sigma$~Orionis cluster 
had not been brought up yet.} 
\item[$^{d}$] {There is independent spectroscopic confirmation of cluster
non-membership of the star No.~33, for which Sacco et~al. (2008) measured
lithium, radial velocity, low gravity, and H$\alpha$ features inconsistent with
membership.} 
\end{list}
   \end{table*}

\end{document}